# Novel Fabrication of Micromechanical Oscillators with Nanoscale Sensitivity at Room Temperature[*]

Michelle D. Chabot, John M. Moreland, Lan Gao, Sy-Hwang Liou, and Casey W. Miller

*Abstract*—**We report on the design, fabrication, and implementation of ultrasensitive micromechanical oscillators. Our ultrathin single-crystal silicon cantilevers with integrated magnetic structures are the first of their kind: They are fabricated using a novel high-yield process in which magnetic film patterning and deposition are combined with cantilever fabrication. These novel devices have been developed for use as cantilever magnetometers and as force sensors in nuclear magnetic resonance force microscopy (MRFM). These two applications have achieved nanometer-scale resolution using the cantilevers described in this work. Current magnetic moment sensitivity achieved for the devices, when used as magnetometers, is $10^{-15}$ J/T at room temperature, which is more than a 1000 fold improvement in sensitivity, compared to conventional magnetometers. Current room temperature force sensitivity of MRFM cantilevers is ~$10^{-16}$ $N/\sqrt{Hz}$, which is comparable to the room temperature sensitivities of similar devices of its type. Finite element modeling was used to improve design parameters, ensure that the devices meet experimental demands, and correlate mode shape with observed results. The photolithographic fabrication process was optimized, yielding an average of ~85% and alignment better than 1 $\mu$m. Post-fabrication focused-ion-beam milling was used to further pattern the integrated magnetic structures when nanometer scale dimensions were required.**

M. D. Chabot was with the National Institute of Standards and Technology, Boulder, CO 80305 USA. She is now with the Department of Physics, University of San Diego, San Diego, CA 92110 USA (phone: 619-260-8865; fax: 619-260-6874; e-mail: mchabot@sandiego.edu).

J. M. Moreland is with the National Institute of Standards and Technology, Boulder, CO 80305 USA (e-mail: moreland@boulder.nist.gov).

L. Gao and S. H. Liou are with the Department of Physics, University of Nebraska, Lincoln, NE 68588 USA.

C. W. Miller was with the University of Texas, Austin, TX 78723 USA. He is now with the Department of Physics, University of California San Diego, La Jolla, CA 92093 USA.

* Work partially supported by US Government, not subject to US Copyright.

# I. INTRODUCTION

OVER the past decade, several experimental methods have been developed to probe material properties on the micrometer and nanometer scales[1]-[7]. Many of these novel methods employ the use of micromechanical cantilevers to achieve the desired sensitivity. Because these experiments are limited by the thermal noise of the cantilever itself, low temperatures often must be used, making the results less relevant to industrial applications. Thus, there is a strong demand for microcantilevers that are sensitive enough to obtain useful results at room temperature. A further complication arises when the experiments require that a micrometer-sized magnetic material be placed onto the cantilever. Doing this on an individual basis is not only time-consuming, but it also jeopardizes the uniformity and consistency of the results. Therefore, there is a clear need to develop a process to batch-fabricate ultrasensitive cantilevers with magnetic dots pre-aligned and deposited as part of the microfabrication process. In this article, we describe a process that has been developed to meet these demands. Careful consideration was given to the design of the oscillator shape, and finite-element modeling was used to study the resonant shapes and to make sure that the resonance frequencies were in the desired ranges for the specific applications.

The fabrication process involved double-sided alignments and multiple exposures, with both wet and dry etches used at different processing steps. One wafer produces 30 chips, each connected to a frame by easy-break tabs. Each chip has 10 devices, giving an ideal yield of 300 devices/wafer. The actual process yield is approximately 85%, resulting in ~250 devices per wafer. The devices have been successfully used as both micromagnetometers and as force sensors in nuclear magnetic resonance force microscopy (MRFM). We report the results of these experiments, showing that the observed nanometer scale sensitivity correlates with the predicted sensitivities from modeling.

## II. DEVICE DESIGN

*A. Overview*

These devices have been specifically designed with magnetometry and MRFM in mind. Each application has separate requirements and must be examined individually. A fabrication process has been developed that is flexible enough to accommodate the various demands. In order to fully understand the device design, the applications and demands of magnetometry and MRFM are detailed below.

*B. Design Considerations: Magnetometry*

Microcantilever magnetometry is a novel method that has been developed to make micrometer and nanometer scale magnetic measurements, a feat that has proven to be a challenge for conventional magnetometers[8]-[14]. Fig. 1 shows the basic setup for this measurement technique. A thin magnetic film is placed on a torsional cantilever, and an external magnetic field, $H_0$, is applied in the cantilever plane. A small torque field $H_T$ is applied perpendicular to the cantilever plane and is oscillated at the cantilever resonant frequency. The interaction of $H_T$ with the sample's magnetization results in an oscillating torque $\tau$, which resonantly drives the cantilever. The amplitude of oscillation is directly related to the torque, and therefore the magnetization. Magnetization is then obtained as a function of swept $H_0$.

The sensitivity of microcantilever magnetometry is limited by the thermal noise of the cantilever. This can be expressed in terms of the minimum torque required to create an observable signal, which is given by

$$\tau_{min} = \sqrt{\frac{4 k_B T \kappa}{Q \omega_0}} \quad \frac{\text{N} \cdot \text{m}}{\sqrt{\text{Hz}}}, \quad (1)$$

where $\kappa$ is the torsional spring constant, $Q$ is the quality factor, $\omega_0$ is the resonant frequency of the cantilever, $T$ is the temperature, and $k_B$ is Boltzmann's constant. Thus, in order to achieve nanometer-scale resolution at room temperature, devices must be fabricated that have a high quality factor $Q$, high resonant frequency $\omega_0$, and low spring constant $\kappa$. The torsional spring constant for a bar twisting about an axis running through its middle and along its length is given by

$$\kappa = \frac{E\,w\,t^3}{6\,l\,(1+n)}, \qquad (2)$$

where $E$ is Young's modulus, $n$ is the Poisson ratio, $w$ is the bar width, $l$ is the length, and $t$ is the thickness.

As is evident from Eq. (2), the best way to reduce the spring constant (and therefore increase sensitivity) is to decrease the cantilever thickness. However, due to practical limitations, the thickness of the cantilevers must be kept above 150 nm to provide a sufficiently sturdy design. Additionally, increasing the resonant frequency also increases sensitivity, but cantilever magnetometry requires that the resonant frequency be kept below approximately 200 kHz in order to accommodate the ac magnetic field $H_T$ used in the excitation.

Since, in this case, the magnetic structures are the samples, the shape of the structures for this application depends on the desired study. Additionally, because repeatable results for an investigation of shape effects are desired, uniformity and consistency between devices is crucial for this application. This homogeneity between devices will allow for multiple runs on different devices all with magnetic structures of the same shape, thus allowing for this technique to rule out effects caused by material defects.

*C. Design Considerations: Force Detection of Nuclear Magnetic Resonance (NMR)*

Another application for these ultra-sensitive devices is magnetic resonance force microscopy[1], [15]-[19]. This is a novel technique for force detection of electron spin, nuclear magnetic, or ferromagnetic resonance. Henceforth we will focus solely on the force detection of nuclear magnetic resonance(NMR). Figure 2 shows the specific MRFM set-up for which these cantilevers are designed. A sample is placed close to a microcantilever that is in a large polarizing, homogenous external magnetic field. A magnetic dot on the cantilever is saturated by the external field and serves to locally perturb the homogenous field, resulting in a field gradient in the region of the sample. A frequency modulated rf field is used to adiabatically invert the spins residing within a "resonant slice" of the sample. With this cyclic adiabatic inversion method, the frequency modulation parameters are used to invert the sample magnetization within the resonant slice at a frequency that matches the mechanical resonant frequency of the cantilever. This oscillating magnetization in the field gradient of the magnetic dot results in an oscillating force that drives the cantilever into resonance. The amplitude of oscillation (typically detected with a laser interferometer), quality factor, and spring constant of the cantilever yield the force on the cantilever due to the moments in the resonant slice of the sample.

This technique is limited by the thermal noise of the cantilever, which can be expressed as $F_{min} = \sqrt{\frac{4 k_B T k}{Q \omega_0}} \ \frac{N}{\sqrt{Hz}}$, where the spring constant for a bar bending about one end is given by:

$$k = \frac{E w t^3}{4 l^3} \ . \qquad (3)$$

The practical limitations for this technique put restrictions on the cantilever specifications that are different from the magnetometry previously discussed. The main difference is that the mechanical resonance frequency must be kept below $\approx 15$ kHz in order to be able to use cyclic adiabatic inversion to manipulate the nuclear moments. A sturdy design is still required, so again a minimum thickness of 150 nm is desired.

The magnetic dot on the cantilever is made cylindrical in shape to allow for a straightforward calculation of the magnetic-field gradient (for example, see Fig. 3 and Fig. 6). As with most magnetic resonance techniques, higher field gradients mean better resolution. This would dictate thin films of small diameter. However, in order to image deeply within a sample the magnetic dot needs to be as thick as possible so that there is a large field gradient far away from the surface of the dot. For the process described here, the magnets have been made with thicknesses up to 370 nm and diameters of 3 – 5 μm, which is sufficient to provide nanometer-scale resonance slices several micrometers from the magnet.

*D. Finite Element Modeling and Device Geometry*

Both microcantilever magnetometry and MRFM benefit from cantilevers of high-$Q$ and low spring constant. The differing requirements for the resonance frequency are handled by changing shape and lateral dimensions. Finite-element modeling was used to find the cantilever geometry that would best accommodate specific resonant frequencies and a low spring constant. As already mentioned, the most efficient way to achieve a low spring constant is to decrease the thickness, yet the thickness is directly related to the final resonant frequency. Before fabrication begins, resonant frequencies and mode shapes were examined as a function of thickness for a given geometry. From the results, the range of acceptable final device thickness was determined. Typically, devices intended for use in MRFM have allowable thicknesses from 150 nm to 350 nm. Above 350 nm, the resonant frequencies for the specific geometries exceed the 15 kHz limit. For devices intended for use as magnetometers, the range is much larger, generally from 150 nm to 800 nm.

In addition to improving sensitivity by decreasing thickness, making the aspect ratio ($l/w$) as large as possible, without sacrificing stability, is also beneficial. This effect is clear from examining equations (2) and (3). With this in mind, two main design shapes have been fabricated to accommodate the different applications. The first is the simple bar shown in Figure 3. These more traditional devices are intended for use in their lower bending mode. The larger paddle area is made to be at least 30 μm on a side to serve as a platform for the measurement of the oscillation amplitude, which is generally made using fiber-optic laser interferometry. A typical bar cantilever from our processing has a thickness of 200 nm, a width of 3 μm, and a length of 180 μm. These values result in a predicted bending spring constant of $1.3 \times 10^{-4}$ N/m. Finite-element modeling predicts the resonance frequency for the lowest mode to be approximately 10 kHz, which is well within the desired range for employing cyclic adiabatic inversion of the nuclear spins in MRFM.

A second design is shown in Figure 4 and this design is most applicable to microcantilever magnetometry because of the existence of multiple mode shapes [20-22]. These double torsional oscillators consist of a small head connected to a larger wing, which is attached to a fixed base. The necks joining the head, wing, and base are free to twist or bend. This design results in four main modes of operation: the lower and upper bending modes and the lower and upper torsional modes. Figure 5 shows snapshots from the finite-element modeling results for these four modes. The lighter color in the figure indicates the largest amount of strain. From the contour map of the strain in each mode the upper torsional mode has clearly little to no strain that is directly coupling to the fixed base. This allows for a decrease in the damping due to energy lost to the base, and therefore an increase in the $Q$ of the cantilever.

The benefits of this geometry are twofold: not only is the $Q$ increased when working in the upper torsional mode, but the resonance frequency of the upper torsional mode is generally higher by an order of magnitude than that for the lower modes. These two factors combine to make the sensitivity of the upper torsional mode better by almost an order of magnitude than that for the corresponding lower modes. The ability to increase sensitivity through geometric considerations alone is a major factor that has allowed for the development of ultrasensitive cantilevers that can operate at room temperature.

The one weakness of the torsional geometry is the decrease that occurs in the spring constant because the aspect ratio of the neck is about one tenth of what it is for the bar geometry. More complex geometries have been created that increase the aspect ratio of the neck without significantly affecting the other properties. Two such double torsional oscillators are shown in Figure 6.

*E. Predicted sensitivities*

For a thickness of 300 nm, the upper torsional mode of the cantilever shown in Fig. 4 is expected to have a torsional spring constant of $1 \times 10^{-10}$ N·m and a resonant frequency of 120 kHz. A typical value for the $Q$ of the upper torsional mode at room temperature and a moderate vacuum of 13 Pa is 12,000. This corresponds to a minimum detectable torque of $1.3 \times 10^{-20}$ N·m/$\sqrt{Hz}$. In microcantilever magnetometry, this results in a single-sweep magnetic-moment sensitivity at room temperature of $10^{-15}$ J/T (~$10^8$ $\mu_B$). For $Ni_{80}Fe_{20}$ films, this is equivalent to a cubic structure 130 nm on a side, which is well below the sensitivity of any conventional magnetometer.

For a thickness of 200 nm, the lower bending mode of the cantilever shown in Fig. 3 is expected to have a spring constant of $1.3 \times 10^{-4}$ N/m and a resonant frequency of 10 kHz. A typical $Q$ for the bar geometry at moderate vacuum is ~5000. This corresponds to a minimum detectable force of $1.4 \times 10^{-16}$ N/$\sqrt{Hz}$. In a standard force detection of NMR experiment, this results in the ability to image a sample slice 4 μm in diameter and 200 nm thick, with an expected single-shot signal-to-noise ratio of 5 at room temperature.

## III. DEVICE FABRICATION

*A. Process outline*

The fabrication process was designed to allow for high yield and intricate patterning. Fig. 7 shows the process outline. The starting wafer was (100) silicon with a 2 μm boron layer diffused into the top side. The boron diffusion was done at 1150 °C by use of a boron nitride high-temperature planar-diffusion source. After the diffusion, the boron skin was removed by means of a quick HF soak followed by a short reactive ion etch (RIE). Then 1 μm of nitride was deposited onto both sides, and the bottom was patterned using photolithography. The exposed nitride on the bottom was then etched in the RIE to 0.5 μm. A second mask was aligned to the bottom and the wafer patterned. The exposed nitride on the bottom was etched using the RIE by an additional 0.5 μm, so that in some areas the nitride had been completely removed and bare silicon was exposed.

At this point the wafer was placed in a KOH solution, with the final goal of this back-etching step to be the creation of boron-doped silicon membranes. The solution consisted of 35 % KOH and 65 % water kept at a constant temperature of 85 °C. The high KOH concentration and the high temperature result in the most uniform etch, which is an important requirement for this process. The wafer was periodically rotated by 90° to insure a uniform etch. It was kept in the etching solution for 3 h, until approximately ¾ of the total wafer thickness had been etched through. At this point, the wafer was removed, and a 0.5 μm RIE was done on the bottom to remove the nitride from the areas that will form the easy-break tabs that connect each chip to the main silicon frame. The wafer was then placed back in the 35 % KOH solution for 1 hour until light shining from the front appeared red-colored through the etched areas. This reddish tint indicates that the remaining silicon membrane is approximately 15 μm thick. The wafer was removed from the solution at this time and placed in a 10 % KOH, 90 % water solution at 70 °C. It is known that the lower concentration solutions stop KOH etching more effectively on highly hole-doped silicon [23]-[24]. For a 10 % solution, the etch rate decreases by a factor of 100 for concentrations above $10^{20}$ ions/cm$^3$. The lower temperature simply slows the total etch rate down so that the wafer can be carefully removed at the appropriate time. The wafer was removed when light shining from the front became a uniform yellow over the entire membrane; the wafer was then cleaned.

The resulting membranes consisted of 2 µm of boron-doped silicon protected on the front by 1 µm of nitride. The wafer was then placed in the RIE and the top 1 µm of nitride was completely removed, leaving the boron-doped silicon exposed from both sides. Next, in order to integrate small magnetic structures, a photolithography mask was back-aligned to marks on the bottom of the wafer. The wafer was exposed and developed using a two-resist process. A magnetic film was deposited, and a subsequent lift-off leaves only the desired patterned structures on the thin boron membrane. This third mask and the corresponding deposition is what determines the shape and thickness of the magnetic dots. Dots as thin as 15 nm and as thick as 370 nm have been made by adjusting the lift-off resist used.

A fourth and final photolithography step was used to align and pattern the cantilevers. The cantilevers were patterned on the membranes in photoresist with <1 µm alignment, and an RIE was performed to etch away the single-crystal silicon boron-doped membrane areas that were not protected by the photoresist. This final RIE can be carefully timed so that the devices are released before the photoresist is completely etched away. When this is the case, the RIE is continued until all the photoresist is gone, and the wafer is immediately removed from the RIE. When the photoresist is etched away before the devices are released, the magnetic structures are positioned on silicon pillars that are formed during this last etch step, as can be seen clearly in Fig. 10(b). Either way, this final RIE determines the final thickness of the cantilevers. This can be easily controlled by performing an initial back etch of the boron membrane before patterning in order to reduce the starting thickness. This optional step, along with specific RIE parameters and photoresist thickness, allows for control of the thicknesses for ranges from 100 to 1500 nm. Devices as thin as 150 nm have been achieved, as determined by side-view scanning electron micrographs (Fig. 8). Additionally, note that this final RIE release does not destroy or contaminate the magnetic structures. The average yield for this process is 85 %.

## B. Integrated micro- and nanometer size magnetic structures

Permalloy ($Ni_{80}Fe_{20}$) was chosen as the magnetic material for the microstructures because of the low amount of oxidation that occurs in an ambient environment. If a more reactive magnetic material is desired, a simple capping layer can be deposited to solve this problem. The Permalloy films were prepared by thermal evaporation at a pressure of $1.2 \times 10^{-4}$ Pa and an evaporation rate of 0.5 nm/s.

Fig. 9 shows several devices patterned and fabricated using the above process[25]. Single structures as small as 1 µm have been consistently aligned and patterned during batch fabrication. In addition, several devices have been fabricated that have arrays of 1 µm structures with size and spacing varying by less than 10 % (Fig. 9b). Adjacent devices can be patterned with differing structures in order to obtain results that can be reliably compared to each other (Fig. 9c,d).

The film patterning in the batch process is limited by photolithography, and structures smaller than 1 µm have been obtained using focused ion beam milling as a post-fabrication step. Figure 10 shows three devices that have been patterned with interesting nanometer-scale shapes that can be studied using microcantilever magnetometry[25].

## IV. SENSITIVITY DEMONSTRATIONS

### A. Magnetometry

Microcantilever magnetometry offers the ability to investigate micrometer and nanometer scale magnetism on individual structures. Many noncantilever-based measurements are currently made on arrays of micromagnetic dots[7], [26], [27]. However, these results are clouded due to statistical variations within the array such as dot shape, size, and spacing. Efforts are underway to improve fabrication techniques to minimize these effects, which are an integral and unavoidable part of any array measurements. Furthermore, adjacent dots interact magnetostatically. Therefore microcantilever magnetometry offers the great advantage of being able to investigate the properties of single structures. This method hinges on being able to obtain well-defined structures on ultrasensitive cantilevers; the processing described above has overcome this challenge.

The micromechanical cantilevers with integrated samples have been successfully implemented as magnetometers. A double torsional cantilever was used, as illustrated in Fig. 1. The torsional spring constant was calculated to be $5 \times 10^{-9}$ N m. The resonant frequency of the upper torsional mode was 120 kHz with a $Q$ of 12,000 at 13 Pa. For the lower torsional mode, the resonant frequency was 50 kHz with a $Q$ of 4000. A 5 µm × 5 µm × 30 nm (total volume of $7.5 \times 10^{-19}$ m$^3$) Ni$_{80}$Fe$_{20}$ film was patterned onto the head, as shown in Fig. 4(b). The magnetization vs. external field hysteresis loop is shown in Fig. 11. The measured torque at saturation was $4.2 \times 10^{-17}$ N·m. The signal-to-noise ratio was $50 \pm 5$, indicating a minimum detectable torque of $8.4 \pm 0.7 \times 10^{-19}$ N·m. This corresponds to a magnetic-moment sensitivity of $6.7 \times 10^{-15}$ J/T ($7.2 \times 10^{8}$ µ$_B$). This is within 15 % of the predicted value of $7.1 \times 10^{-19}$ N·m, calculated using a lock-in amplifier bandwidth $\Delta v = (4 \times 30 \text{ ms})^{-1}$. These results show that the devices can be reliably implemented as magnetometers with nanometer-scale sensitivity at room temperature.

## B. Magnetic Resonance Force Microscopy

In addition to being successfully used as ultrasensitive magnetometers, the cantilevers have been used as force sensors in a magnet-on-oscillator demonstration of MRFM [28, 29]. In this experiment, a bar oscillator was used as shown in the top image of Fig. 3. A cylindrical Permalloy ferromagnet 4 µm in diameter and 170 nm thick was used. The resonance frequency of the lower bending mode was found to be 4.0 kHz. The room-temperature experiment was performed in an exchange gas pressure of $\sim 10^{-4}$ Torr, resulting in a pressure-limited $Q$ of 1600. The external polarizing field was 8.1 T. The spring constant was measured to be $4 \times 10^{-4}$ N/m. These parameters corresponded to a predicted force sensitivity of $4.1 \times 10^{-16}$ $N/\sqrt{Hz}$.

The experiment measured proton ($^1$H) NMR from a large ($\sim 1$ mm$^3$), ammonium sulfate crystal representing a semi-infinite slab. Ammonium sulfate was used for its long room temperature spin-lattice relaxation time ($\sim 5$ s), large proton density (6.5 x $10^{22}$ /cm$^3$), and easy cleavability; the latter results in a sharp sample-vacuum interface. The sample was placed a few micrometers from the cantilever. The position of the resonant slice was shifted in 500 nm steps by changing the carrier frequency of the rf field. Cyclic adiabatic inversion was performed four times for each carrier frequency for a duration of 1.8 s, embedded in a total rf exposure of 2.2 s. The decay to resonance had a time constant of 10 ms. Frequency modulation with an amplitude of $\sim(50$ kHz$)(2\pi)$ was initiated 10 ms after the decay to resonance. The rf field ($H_1$) was estimated to be $\sim 5$ G.

Figure 12 shows the NMR-induced signal as a function of resonance slice position. A smooth background signal artifact due to induced oscillator motion from the frequency modulated rf field has been subtracted from the raw data. The RMS value of the resultant oscillator displacement was 3.9 nm, and the RMS noise level was 1.0 nm, corresponding to forces of $9.7 \times 10^{-16}$ N and $2.5 \times 10^{-16}$ N, respectively. These results are in reasonable agreement with the expected signal of $7.6 \times 10^{-16}$ N. These data show a sharp step at $\sim 2$ µm; before this point the resonant slice is outside of the sample (close to the Permalloy dot on the oscillator), while beyond this point, the resonant slice is immersed within the sample. The resonant slice was estimated to be approximately 200 nm thick near the sample-vacuum interface from modeling the field gradient of the dot. This experiment demonstrates that the cantilevers with integrated magnetic structures can be successfully implemented as force sensors with nanometer-scale sensitivity at room temperature.

## V. CONCLUSION

These novel micromechanical structures have already had a strong impact on several nanoscale measurement techniques. The microcantilever magnetometry experiments have expanded the capabilities of research techniques in the field of micro- and nanomagnetism. These cantilevers have the sensitivity to measure the quantitative hysteresis loop of an individual single-domain structure, a capability that has not been obtained by any other measurement technique to date.

As with all micromechanical oscillators, the thermal noise will decrease significantly at lower temperatures due to the dependence on $T$ and because the $Q$ increases significantly with temperature. Therefore, even greater sensitivity (~100 to ~1000 better) can be achieved if these devices are used in low-temperature applications.

## ACKNOWLEDGMENT

Support for C. W. Miller while at University of Texas was provided by Army Research Office Contract No. DAAD-19-02-C-0064.

## REFERENCES

[1] D. Rugar, B. C, Stipe, H. J. Mamin, C. S. Yannoni, T. D. Stowe, K. Y. Yasumura, and T. W. Kenny, "Adventures in attonewton force detection," *Appl. Phys. A*, vol. 72[Suppl.], pp. S3-S10, March 2001.

[2] W. Wernsdorfer, D. Mailly, and A. Benoit, "Single nanoparticle measurement techniques," *J. Appl. Phys.*, vol. 87, pp. 5094 -5096, May 2000.

[3] R. P. Cowburn, D. K. Koltsov, A. O. Adeyeye, and M. E. Welland, "Probing submicron nanomagnets by magneto-optics," *Appl. Phys. Lett.*, vol. 73, pp. 3947 – 3949, Dec. 1998.

[4] R. H. Kock, J. G. Deak, D. W. Abraham, P. L. Trouilloud, R. A. Altman, Yu Lu, W. J. Gallagher, R. E. Scheuerlein, K. P. Roche, and S. S. P. Parkin, "Magnetization reversal in micron-sized magnetic thin films," *Phys. Rev. Lett.*, vol. 81, pp. 4512 – 4515, Nov. 1998.

[5] M. Lederman, S. Schultz, and M. Ozaki, "Measurement of the Dynamics of the Magnetization Reversal in Individual Single-Domain Ferromagnetic Particles," *Phys. Rev. Lett.*, vol. 73, pp. 1986 – 1989, Oct. 1994.

[6] S. E. Russek, S. Kaka, and M. Donahue, "High-speed dynamics, damping, and relaxation times in submicrometer spin-valve devices," *J. Appl. Phys.*, vol. 87, pp. 7070 – 7072, May 2000.

[7] U. Wiedwald, M. Spasova, M. Farle, M. Hilgendorff, and M. Giersig, "Ferromagnetic resonance of monodisperse Co particles," *J. Vac. Sci. Technol. A*, vol. 19, pp. 1773 – 1776, Aug. 2001.

[8] M. D. Chabot and J. Moreland, "Micrometer-scale magnetometry of thin Ni-Fe films using ultra-sensitive microcantilevers," *J. Appl. Phys.*, vol. 93, pp. 7897 – 7899, May 2003.

[9] B. C. Stipe, H. J. Mamin, T. D. Stowe, T. W. Kenny, and D. Rugar, "Magnetic dissipation and fluctuations in individual nanomagnets measures by ultrasensitive cantilever magnetometry," *Phys. Rev. Lett.*, vol. 86, pp. 2874 – 2877, March 2001.

[10] J. G. E. Harris, D. D. Awschalom, F. Matsukura, H. Ohno, K. D. Maranowski, and A. C. Gossard, "Integrated micromechanical cantilever magnetometry of GaMnAs," *Appl. Phys. Lett.*, vol. 75, pp. 1140 – 1142, Aug. 1999.

[11] M. Löhndorf, J. Moreland, P. Kabos, and N. Rizzo, "Microcantilever torque magnetometry of thin magnetic films," *J. Appl. Phys.*, vol 87, pp. 5995 – 5997, May 2000.

[12] J. G. E. Harris, R. Knobel, K. D. Maranowski, A. C. Gossard, N. Samarth, and D. D. Awschalom, "Magnetization measurements of magnetic two-dimensional electron gases," *Phys. Rev. Lett.*, vol. 86, pp. 4644 – 4647, May 2001.

[13] C. Rossel, P. Bauer, D. Zech, J. Hofer, M. Willemin, and H. Keller, "Active microlevers as miniature torque magnetometers," *J. Appl. Phys.*, vol. 79, pp. 8166 – 8173, June 1996.

[14] C. Lupien, B. Ellman, P. Grutter, and L. Taillefer, "Piezoresistive torque magnetometer below 1 K," *Appl. Phys. Lett.*, vol. 74, pp. 451 – 453, Jan. 1999.

[15] D. Rugar, C. S. Yannoni, and J. A. Sidles, "Mechanical detection of magnetic resonance," *Nature,* vol. 360, pp. 532-566, 1992.

[16] D. Rugar, O. Zuger, S. Hoen, C. S. Yannoni, H. M. Vieth, and R. D. Kendrick, "Force detection of nuclear magnetic resonance," *Science,* vol. 264, pp. 1560-1563, 1994.

[17] K. R. Thurber, L. E. Harrell, and D. D. Smith, "170 nm nuclear magnetic resonance imaging using magnetic resonance force microscopy," *J. Mag. Res.*, vol. 162, pp. 336-340, 2003.

[18]   J. A. Sidles, J. L. Garbini, and G. P. Drobny, "The theory of oscillator-coupled magnetic resonance with potential applications to molecular imaging," *Rev. Sci. Instrum.*, vol. 63, pp. 3881-3899, 1992.

[19]   D. Rugar, R. Budakin, H. J. Mamin, and B. W. Chui, "Single spin detection by magnetic resonance force microscopy," Nature, vol. 430, pp. 329-332, 2004.

[20]   R. N. Kleinman, G. K. Kaminsky, J. D. Reppy, R. Pindak, and D. J. Bishop, "Single-crystal silicon high-Q torsional oscillators," *Rev. Sci. Instrum.*, vol. 56, pp. 2088-2091, 1985.

[21]   M. D. Chabot and J. T. Markert, "Microfabrication of single-crystal silicon multiple torsional oscillators," *Proc. SPIE,* vol. 3875, pp. 104-112, 1999.

[22]   C. L. Spiel, R. O. Pohl, and A. T. Zehnder, "Normal modes of a Si(100) double-paddle oscillator," *Rev. Sci. Instrum.*, vol. 72, pp. 1482 – 1491, Feb. 2001.

[23]   H. Seidel, L. Csepregi, A. Heuberger, and H. Baumgartel, "Anisotropic etching of crystalline silicon in alkaline solutions: Influence of dopants," *J. Electrochem. Soc.*, vol. 137, pp. 3626-3632, Nov. 1990.

[24]   E. D. Palik, J. Wl. Faust, Jr., H. F. Gray, and P. F. Green, "Study of the etch-stop mechanism in silicon," *J. Electochem. Soc.*, vol. 129, pp. 2051-2058, 1982.

[25]   L. Gao, D. Q. Feng, L. Yuan, T. Yokota, R. Sabirianov, S. H. Liou, M. D. Chabot, D. Porpora, and J. Moreland, "A study of magnetic interactions of $Ni_{80}Fe_{20}$ arrays using ultrasensitive microcantilever torque mangnetometry," *J. Appl. Phys.*, vol. 95, pp. 7010-7012, June 2004.

[26]   G. Gubbiotti, L. Albini, G. Carlotti, M. De Crescenzi, E. Di Fabrizio, A. Gerardino, O. Donzelli, F. Nizzoli, H. Koo, and R. D. Gomez, "Finite size effects in patterned magnetic Permalloy films," *J. Appl. Phys.*, vol. 87, pp. 5633-5635, May 2000.

[27]   J. Jorzick, S. O. Demokritov, B. Hillebrands, B. Bartenlian, C. Chappert, D. Decanini, F. Rousseaux, and E. Cambril, "Spin-wave quantization and dynamic coupling in micron-size circular magnetic dots," *Appl. Phys. Lett.*, vol. 75, pp. 3859 – 3861, Dec. 1999.

[28]   C. W. Miller, "Nuclear Magnetic Resonance Force Microscopy: Adiabaticity, External Field Effects, and Demonstration of Magnet-on-Oscillator Detection with Sub-Micron Resolution," *Ph.D. Dissertation for the University of Texas at Austin*, Dec. 2003.

[29]   J.-H Choi, U. M. Mirsaidov, C. W. Miller, Y. J. Lee, S. Guchhait, M. D. Chabot, W. Lu, and J. T. Markert, "Oscillator microfabrication, micromagnets, and magnetic resonance force microscopy", Proc. SPIE Int. Soc. Opt. Eng., vol. 5389, pp. 399-410, July 2004.

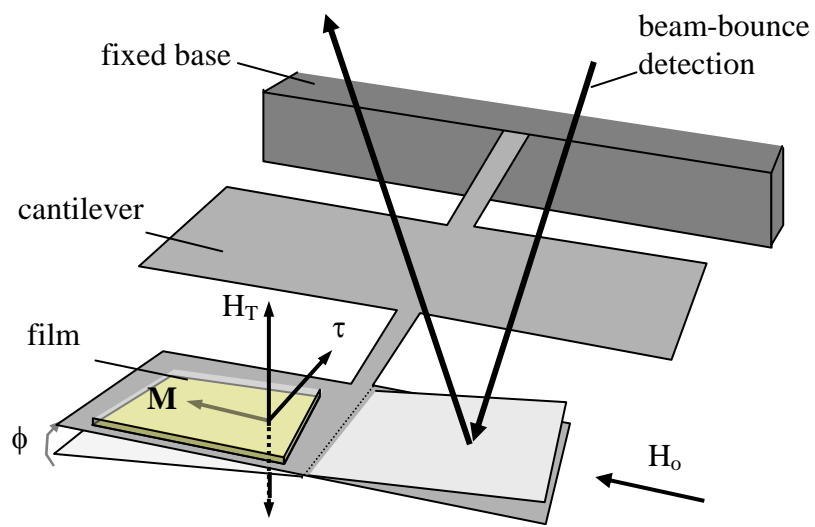

Fig. 1. Overview of microcantilever magnetometry using a double torsional oscillator. In this illustration, the amplitude of oscillation is detected by means of a laser beam-bounce.

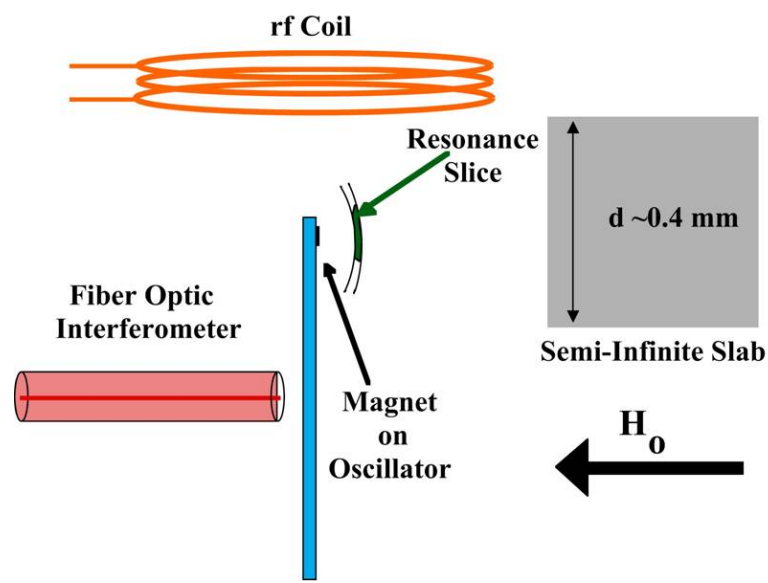

Fig. 2. Overview of nuclear magnetic resonance force microscopy set-up for which these devices were designed. Notice that the external magnetic field lies perpendicular to the cantilever plane.

CHABOT 19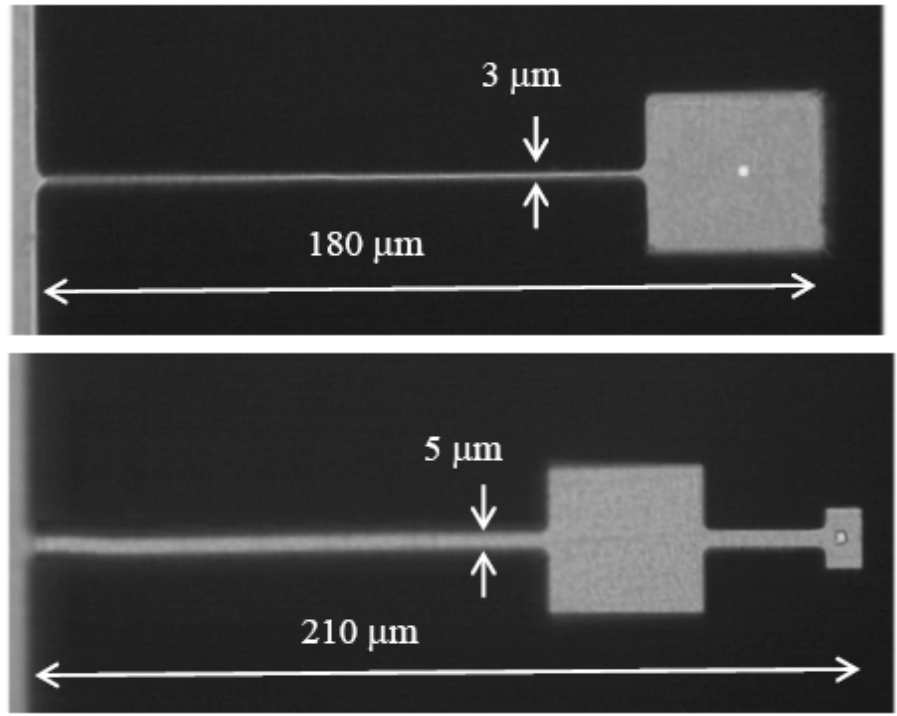

Fig. 3. Optical photographs of two different bar geometries. These cantilevers obtain their nanometer scale sensitivity by having extremely low spring constants because they have an extremely large aspect ratio (~60) and very small thicknesses (~200 nm thick). Each has magnetic film 3 μm in diameter aligned and deposited onto the upper paddle.

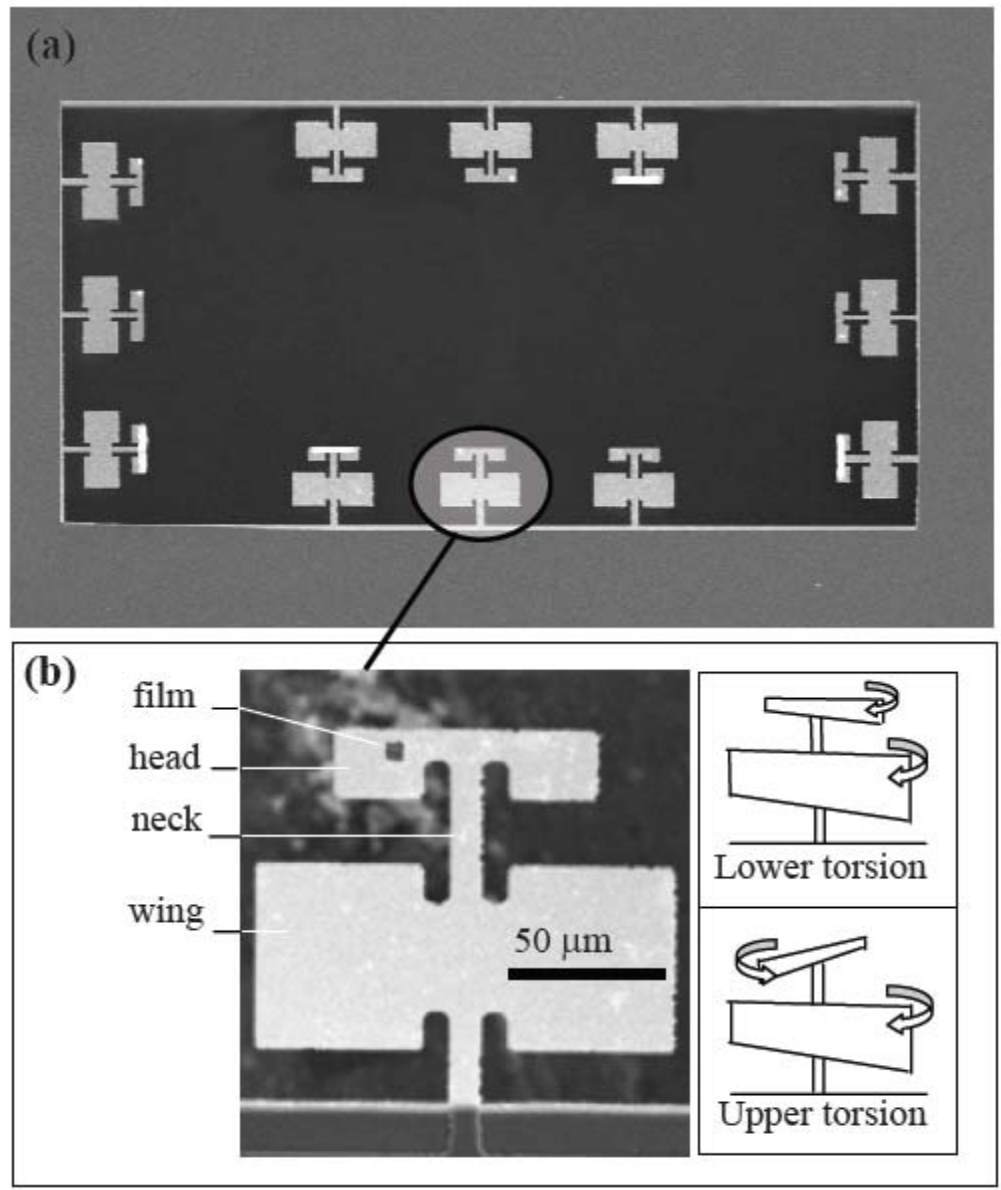

Fig. 4. Scanning electron micrographs (SEMs) of the double torsional oscillator geometry. (a) A typical chip containing 12 devices all with double-side access. (b) A closer view of a double torsional oscillator with a 5 μm × 5 μm × 30 nm film on the head. The illustration on the right side indicates the shape of the two main torsional modes of operation.

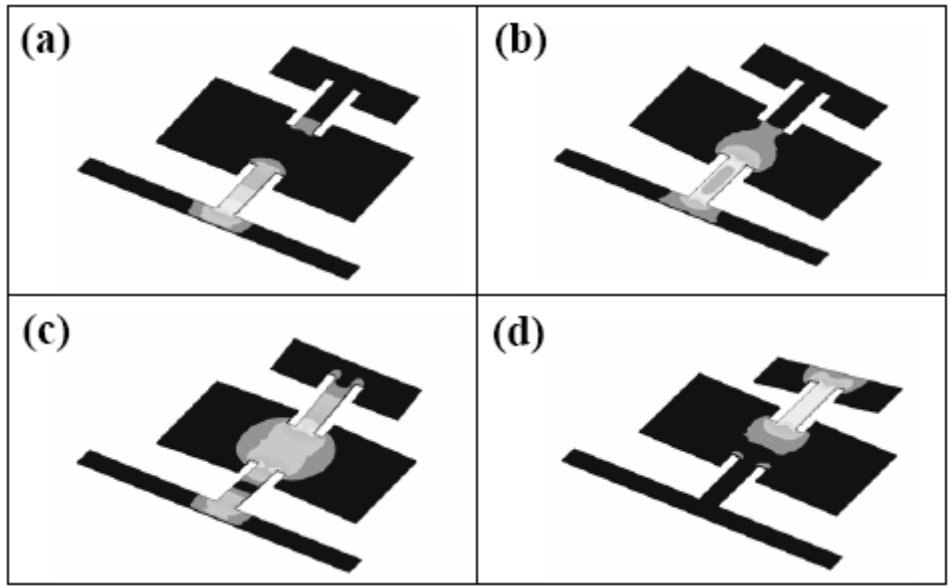

Fig. 5. Finite-element modeling results showing the (a) lower bending mode, (b) lower torsional mode, (c) upper bending mode, and (d) upper torsional mode. Light coloring indicates areas of highest strain.

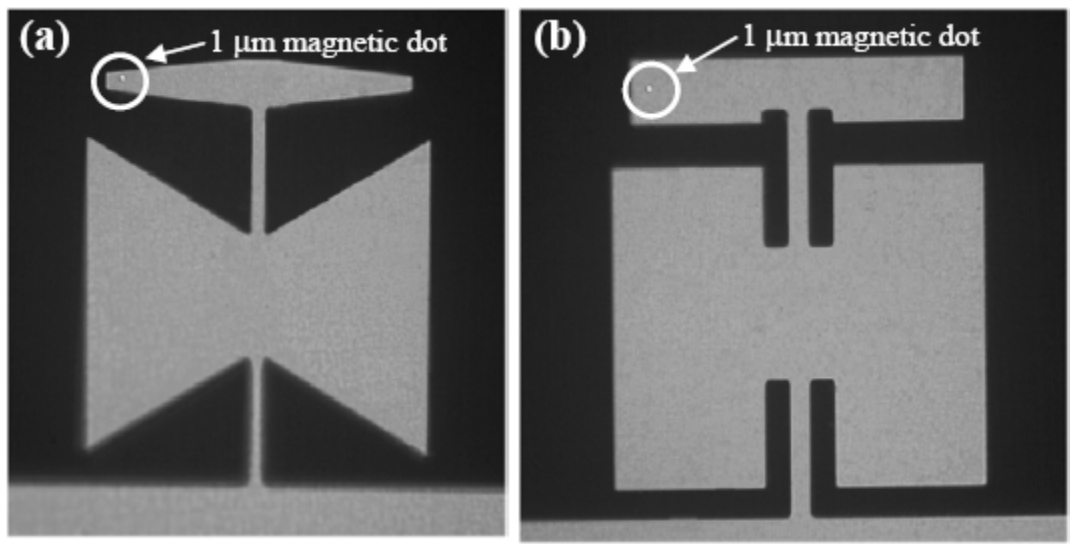

Fig. 6. Optical photographs of two different double torsional geometries. (a) The bowtie cantilever, which reduces the ratio of the mass of the head to the mass of the wing without sacrificing the length of the neck. This geometry further reduces the energy lost to the base. (b) An exaggerated version of the standard geometry allows for necks up to 50 μm long while at the same time maintaining the large ratio between the mass of the wing and the mass of the head.

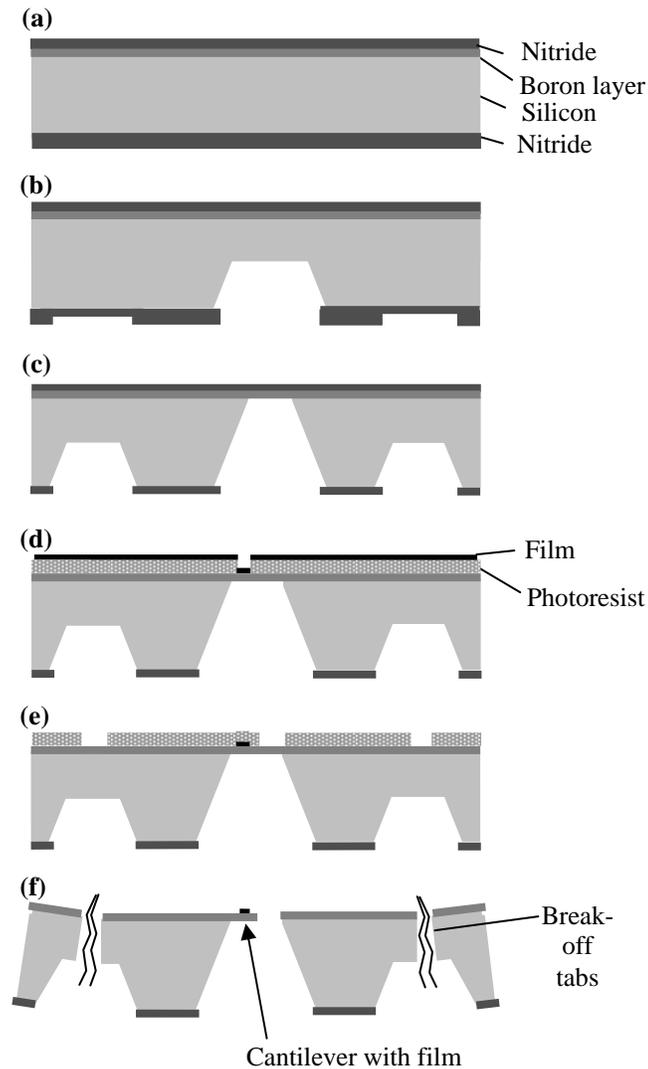

Fig. 7. Fabrication process overview. (a) The starting wafer is (100) silicon with a 2 μm boron layer diffused into one side. Nitride is deposited onto both sides. (b) The bottom is patterned with multiple photolithography masks and the exposed silicon is etched part way with a KOH solution. (c) The bottom nitride is etched to remove the second nitride step, and the wafer is placed back in the KOH until it stops on the boron layer. (d) The top nitride is removed and photoresist is spun and patterned. A magnetic film is deposited. (e) A lift-off is performed and a second layer of photoresist is patterned. (f) A final reactive ion etch releases the cantilevers. Each chip is attached to a silicon main frame by break-off tabs.

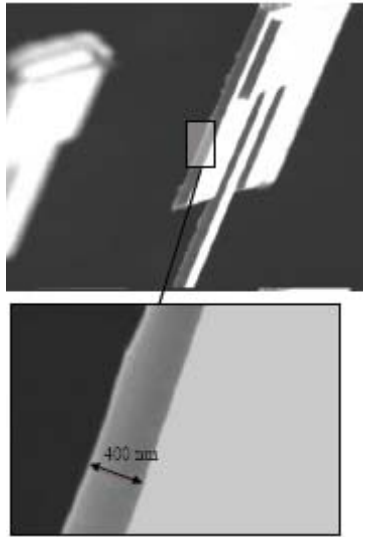

Fig. 8. Side-view scanning electron micrographs (SEMs) are taken to determine device thickness. The double torsional cantilever shown above was found to have a thickness of 400 nm.

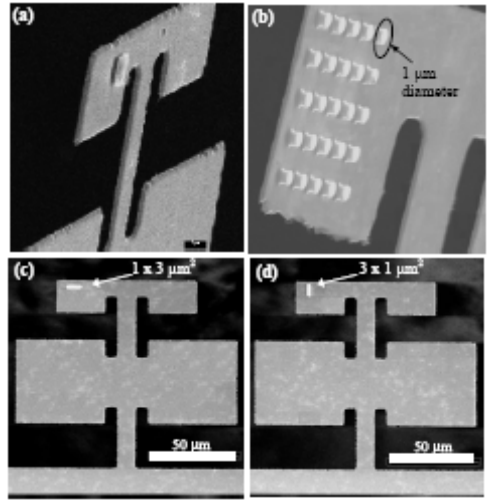

Fig. 9. SEMs of four different double torsional devices intended for use as magnetometers. (a) A 5 μm × 5 μm × 30 nm film is positioned on the head of a cantilever. (b) A 5 × 5 array of 30 nm thick films 1 μm in diameter. (c) and (d) Two adjacent devices are patterned for the study of shape-dependent magnetic switching. The ability to ensure nearly identical experimental set-ups for each shape is crucial in order to obtain reliable results.

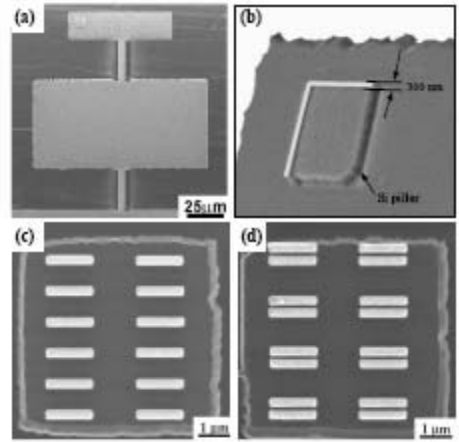

Fig. 10. SEMs of three different double torsional devices intended for use as magnetometers. Each had nanostructures patterned post-fabrication by use of focused ion beam milling[25]. (a) After batch fabrication, the starting cantilever has a 5 μm × 5 μm × 30 nm film positioned on the head. (b) A 300 nm wide × 5 μm long × 30 nm thick L-shaped nanowire on the head of a torsional cantilever. (c) An array of 12 single 300 nm × 32 nm × 1.5 μm bars (d) An array of 8 pairs of 300 nm × 32 nm × 1.5 μm bars.

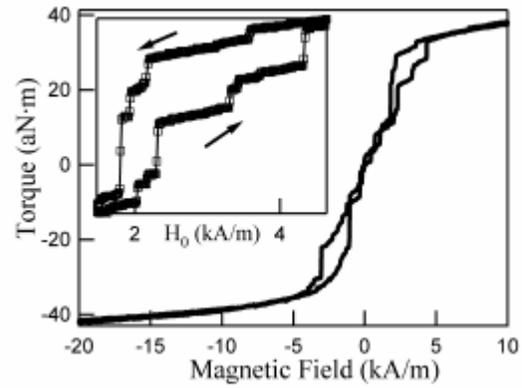

Fig. 11. Torque vs. applied field for a 5 μm ×5 μm × 30 nm Ni-Fe film. Inset: close-up of hysteresis loop showing domain switching.

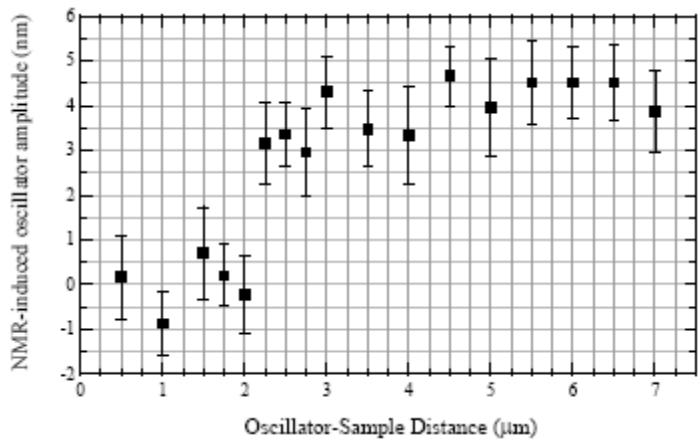

Fig. 12.  NMR-induced oscillator amplitude detected by scanning the carrier frequency to move the resonance slice position.  Each point is the average of four independent measurements.
.